\documentclass[12pt]{iopjournal}
\usepackage{subfigure}
\usepackage{psfrag,graphicx}
\usepackage{bm,color}
\usepackage{latexsym,multirow}
\usepackage{epstopdf}
\usepackage[english]{babel}
\usepackage{amsmath,amssymb,amsfonts}
\usepackage{appendix}
\usepackage{bbold}
\usepackage[dvipsnames]{xcolor}
\usepackage[utf8]{inputenc}   
\usepackage[T1]{fontenc}      
\usepackage{lmodern}          
\usepackage{cite}

\definecolor{mygrey}{gray}{0.35}
\definecolor{myblue}{rgb}{0.2,0.2,0.8}
\definecolor{myzard}{cmyk}{0,0,0.05,0}
\definecolor{mywhite}{rgb}{1,1,1}
\definecolor{mywhite}{rgb}{1,1,1}
\definecolor{myred}{rgb}{1,0.,0.3}

\def\be{ \begin{equation}}
\def\ee{ \end{equation}}
\def\bse{  \begin{subequations}}
\def\ese{  \end{subequations}}
\def\bea#1\ea{\begin{align}#1\end{align}}

\def\bi{\begin{itemize}}
\def\ei{\end{itemize}}

\def\bt{\begin{tabular}}
\def\et{\end{tabular}}
\newcommand{\ket}[1]{|#1\rangle}

\def\sech{\textrm{\,sech\,}}

\begin{document}

\articletype{Paper} 

\title{DRAG-Compatible Leakage Suppression in Landau--Zener Control via Isoprobability Twins}

\author{Ivo S Mihov$^1$ and Nikolay V Vitanov$^2$}

\affil{$^{1,2}$Center for Quantum Technologies, Department of Physics, Sofia University, James Bourchier 5 blvd, 1164 Sofia, Bulgaria}

\keywords{Quantum Control, Quantum Computing}

\begin{abstract}
\vspace{0.2cm}

Analytically solvable models -- particularly the Landau-Majorana-Stückelberg-Zener (LMSZ) and Allen-Eberly-Hioe (AEH) models -- underpin many quantum-gate implementations and population-transfer protocols. However, their canonical pulse shapes are incompatible with modern leakage-suppression techniques and some systems. 
Most notably, the constant Rabi envelope of the LMSZ pulse prevents many leakage-suppresion approaches, which require smoothness. 
We address both limitations by developing the concept of isoprobability twin models: distinct pairs of Rabi frequency $\Omega(t)$ and detuning $\Delta(t)$ that yield identical post-pulse transition probabilities based on the Delos-Thorson transformation.
In this work, we formalise the method by experimentally demonstrating the equivalence of multiple LMSZ and AEH twin models on IBM's \texttt{ibm\_kyiv} processor.
Finally, we show a staggering leakage reduction by more than 3 orders of magnitude using a custom DRAG implementation of a cosine LMSZ isoprobability model.

\end{abstract}


\section{Introduction}

From the scalability of quantum systems to the fidelity of quantum gates, quantum control is key in fuelling the long-sought fault-tolerant era of quantum computing.
Gate implementations usually rely on powerful quantum control approaches carefully crafted for the specific quantum machine.
These approaches include resonant, adiabatic, composite, inverse-engineering or optimal-control techniques that aim to achieve the desired target Hamiltonian.
All these rely on the ability to solve the Schr\"odinger equation accurately, which allows to calculate the propagator that governs the evolution of the quantum state.
It is therefore very beneficial to use quantum models with exact solutions, such as the Landau-Majorana-St\"uckelberg- Zener (LMSZ)~\cite{Landau1932, Majorana1932, Stueckelberg1932, Zener1932, Vitanov1996}, Allen-Eberly-Hioe~(AEH) \cite{Allen1975, Hioe1981}, Rabi~\cite{Rabi1937}, Rosen-Zener (RZ)~\cite{Rosen1932}, Demkov~\cite{Demkov1963}, Demkov-Kunike~\cite{Demkov1969}, Carroll-Hioe~\cite{Hioe1985} model etc.
In addition, many models can be solved approximately -- the Gaussian~\cite{Vasilev2004}, Lorentzian~\cite{Vasilev2014}, Sine~\cite{Yatsenko2004, Boradjiev2013, Mihov20242} models -- resulting in useful analytic expressions.


Bearing important practical applications, such models are fundamental for many quantum computing applications.
For example, a number of common implementations of single-qubit quantum gates are based on LMSZ interferometry~\cite{Ryzhov2024, Hicke2006, Zhang2021, Campbell2020, Caceres2023, Benza2003}.
The LMSZ and AEH models underpin many applications in coherent control well beyond quantum-gate synthesis \cite{Vitanov2001, Vitanov2017}. 
In NMR and MRI, sech/tanh pulses derived from the AEH solution are the basis of $B_1$-insensitive adiabatic spin inversion \cite{Silver1984, Garwood2001}. 
In Rydberg atom physics, chirped-pulse adiabatic rapid passage uses LMSZ-class drives to reach selected Rydberg states with high fidelity, including in the strong-blockade regime \cite{Beterov2011}. 
The broader landscape of driven two-level systems and LZSM interferometry is reviewed in \cite{Shevchenko2010, Ivakhnenko2023}.

We follow a systematic method known as Delos-Thorson equivalence \cite{Vitanov1996, Delos1972, Hioe19852} to identify isoprobability twin models belonging to two of the most widely used analytically solvable classes -- the finite LMSZ class and the AEH class.
Isoprobability twins are models with different Rabi frequency/detuning pairs that produce identical post-pulse transition probability.
In many applications, which depend on the final excited population, rather than full temporal equivalence of the propagator, the approach introduces a new degree of freedom that could be used to increase robustness to errors and leakage and escape hardware limitations.

A standard remedy for noncomputational-state leakage is Derivative Removal by Adiabatic Gate (DRAG)~\cite{Motzoi2009, Gambetta2011, Motzoi2013, Theis2016, Theis2018, Chen2016, Werninghaus2021, Tonchev2025, Patra2026, Aggarwal2026}, which adds an imaginary correction to the Rabi frequency $\Omega(t)$ proportional to $\dot{\Omega}(t)$.
Therefore, it is not applicable to both the original and finite LMSZ pulses, whose constant Rabi envelope results in $\dot{\Omega} = 0$. 
We circumvent this obstruction by introducing isoprobability twin models, as detailed below.


In this work, we use the Delos-Thorson transformation to produce multiple isoprobability twin models belonging to the finite LMSZ class and the AEH class.
We then continue to demonstrate their equivalence on IBM's \texttt{ibm\_kyiv} processor.
Using Rabi frequency phase modulation through the cloud, we evade the fixed detuning limitation of the system~\cite{Dridi2020, Kuzmanovic2024, Kuno2025, Green2015, McCord2025}.
We conclude with a hardware-suited numerical simulation of a new DRAG-corrected LMSZ model that shows a reduction of leakage rates by a factor of 2200.

\section{Delos-Thorson approach}

The Delos-Thorson approach \cite{Vitanov1996, Delos1972, Hioe19852} identifies all pairs $\{\Omega(t), \Delta(t)\}$ that produce the same post-pulse transition probability. The central idea is a change of the independent variable $t$ in the Schr\"odinger equation,
\be\label{SEq}
i \frac{d}{dt} \mathbf{c}(t) = \tfrac{1}{2}
\begin{bmatrix} -\Delta(t) & \Omega(t) \\ \Omega(t) & \Delta(t) \end{bmatrix} \mathbf{c}(t),
\ee
to the Delos-Thorson variable
\be\label{s(x)}
\sigma(t) = \int_0^t \Omega(t') \, dt',
\ee
which recasts the equation of motion as
\be
i \frac{d}{d\sigma} \mathbf{C}(\sigma) = \tfrac{1}{2}
\begin{bmatrix} -\Theta(\sigma) & 1 \\ 1 & \Theta(\sigma) \end{bmatrix} \mathbf{C}(\sigma),
\ee
where the St\"uckelberg variable \cite{Delos1972}
\be\label{Theta(s)}
\Theta(\sigma) = \frac{\Delta(t(\sigma))}{\Omega(t(\sigma))}
\ee
is the sole parameter governing the dynamics. 
Because the transition probability is independent of the choice of integration variable, it depends only on $\Theta(\sigma)$ --- not on $\Omega$ and $\Delta$ individually. 
Consequently, all pairs $\{\Omega(t), \Delta(t)\}$ that generate the same function $\Theta(\sigma)$ share an identical post-pulse transition probability; they form an isoprobability class with generating function $\Theta(\sigma)$. 
Since the pulse shape can be chosen in infinitely many ways, each such class contains infinitely many members that we call isoprobability twins.

To parametrize class members systematically, we write
\be\label{sigma}
\Omega(t) = \Omega_0 f(x), \qquad
\sigma(t) = \Omega_0\tau\, s(x), \qquad
\Delta(t) = \Delta_0 g(x),
\ee
where $x = t/\tau$ is the dimensionless time, $f(x) = ds/dx$, and $s(x) = \int_0^x f(x')\,dx'$. We normalize $f$ so that its temporal area equals $\pi$, giving a pulse area $A = \pi\Omega_0\tau$. New class members can be constructed in two directions:

\subsection*{Given $\Omega(t)$, find $\Delta(t)$}

Pick any pulse envelope $\Omega(t)$ (with the chosen pulse area). 
Integrate it to obtain the Delos-Thorson variable
$\sigma(t) = \int_0^t \Omega(t')\,dt'$, then evaluate the class 
generator $\Theta$ at $\sigma(t)$. From the definition 
$\Theta(\sigma) = \Delta/\Omega$, the paired detuning is
\be\label{eq:Delta}
\Delta(t) = \Omega(t)\,\Theta\bigl(\sigma(t)\bigr).
\ee

\subsection*{Given $\Delta(t)$, find $\Omega(t)$}

Pick any chirp function $\Delta(t)$. Combining $\Theta(\sigma) = \Delta/\Omega$ with $\Omega = \dot\sigma$ yields a separable ordinary differential equation for $\sigma(t)$,
\be\label{eq:ode_for_sigma}
\dot\sigma(t)\,\Theta\bigl(\sigma(t)\bigr) = \Delta(t),
\ee
which integrates by quadrature to
\be\label{eq:sigma_implicit}
\int_0^{\sigma(t)} \Theta(\sigma')\,d\sigma' 
= \int_0^t \Delta(t')\,dt'.
\ee
Inverting this relation gives $\sigma(t)$, and the paired Rabi 
frequency follows as $\Omega(t) = \dot\sigma(t)$.

\section{Testing the equivalence -- isoprobability twins}\label{sec:isoprobability-twins}

Among the dozen exactly soluble analytical models, we choose to examine the transition probability invariance for two classes of models: finite Landau-Majorana- St\"uckelberg-Zener (LMSZ) class and the Allen-Eberly-Hioe (AEH) class.

\subsection{Finite LMSZ class of models}
We consider the finite LMSZ model~\cite{Vitanov1996},
\be\label{LMSZ model}
\Omega(t) = \Omega_0\ \ (|t/\tau|\leqq \pi/2),\quad \Delta(t) = \Delta_0 t/\tau.
\ee
The original LMSZ model features a coupling of infinite duration and an unbounded detuning, making it impossible to implement in an experiment.
The pulse area in the finite LMSZ model is $A=\pi\Omega_0\tau$, the Delos-Thorson variable is 
$\sigma(t) = \Omega_0 t = \Omega_0\tau x $
and $t = \sigma/\Omega_0$.
Hence the St\"uckelberg variable and the phase \eqref{phase} for the LMSZ model read
\be\label{Theta_LMSZ}
\Theta(\sigma) = \frac{\Delta_0}{\Omega_0^2\tau} \sigma, \quad
\varphi(\sigma) = \frac{\Delta_0}{2\Omega_0^2\tau} \sigma^2.
\ee
The transition probability for the finite LMSZ model can be found in~\cite{Vitanov1996}.
For sufficiently long interaction duration it approaches the transition probability for the original LMSZ model \cite{Landau1932, Majorana1932, Stueckelberg1932, Zener1932},
\be
P_{\text{LMSZ}} = 1-\exp \left(-\frac{\pi \Omega_0^2 \tau}{2\Delta_0}\right).
\label{eq-lmsz-original}
\ee
It approaches 1 as $\Omega_0$ increases, which is a common feature of level crossing models in the adiabatic regime.

Let us take now another pulse shape $f(x)$ with the same pulse area as the original model of equation~\eqref{LMSZ model},
\be
f(x) = \frac{\pi}{2} \cos x. 
\ee
Then $s(x) = \frac{\pi}{2} \sin x$ and $\sigma(t) = \frac{\pi}{2} \Omega_0 \tau \sin x$.
We replace this expression in equations~\eqref{eq:Delta} and \eqref{Theta_LMSZ} to find
\bse
\begin{align}
\Omega(t) &= \frac{\pi}{2} \Omega_0 \cos (t/\tau), \quad
\Delta(t) = \frac{\pi^2}{8} \Delta_0 \sin (2t/\tau), \\
\varphi(t) &= \frac{\pi^2}{8} \Delta_0\tau \sin^2 (t/\tau) \quad
(|t/\tau|\leqq \pi/2) .
\end{align}
\ese
This model generates the same transition probability as the original finite LMSZ model \eqref{LMSZ model}.
Furthermore, its Rabi frequency shape is smooth, which will allow us to apply DRAG corrections.

A third choice for a pulse shape is the hyperbolic secant $f(x)=\sech{(x)}$. 
It produces the pair
\bse
\begin{align}
\Omega(t) &= \Omega_0 \sech (t/\tau), \\
\Delta(t) &=  \Delta_0 \sech (t/\tau) \arctan(\sinh(t/\tau)), \\
\varphi(t) &= \frac{1}{2} \Delta_0\tau \arctan^2(\sinh(t/\tau)) \quad
(|t/\tau|\leqq \pi/2) .
\end{align}
\ese

These three pairs are indexed in table~\ref{tab-models} under No. 1, 4 and 8 among a total of 16 pairs $\{\Omega(t),\Delta(t)\}$ that belong to the LMSZ class of models.

\begin{figure}
\begin{tabular}{c}
    \includegraphics[width=0.85\columnwidth]{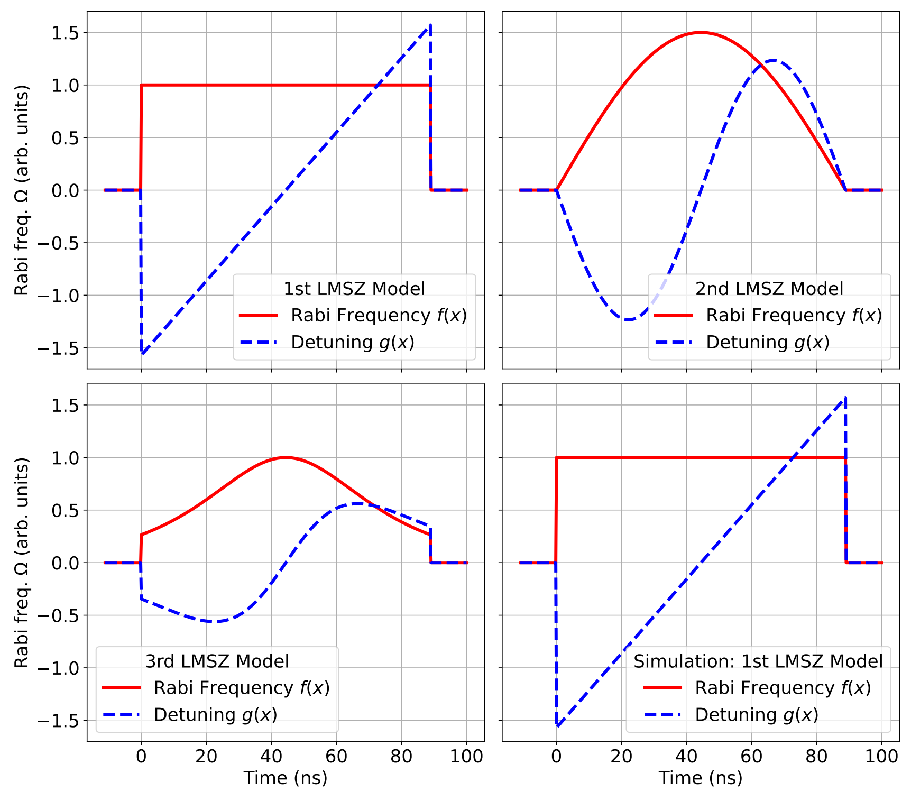} \\
    \hspace*{0.8cm}\includegraphics[width=0.85\columnwidth]{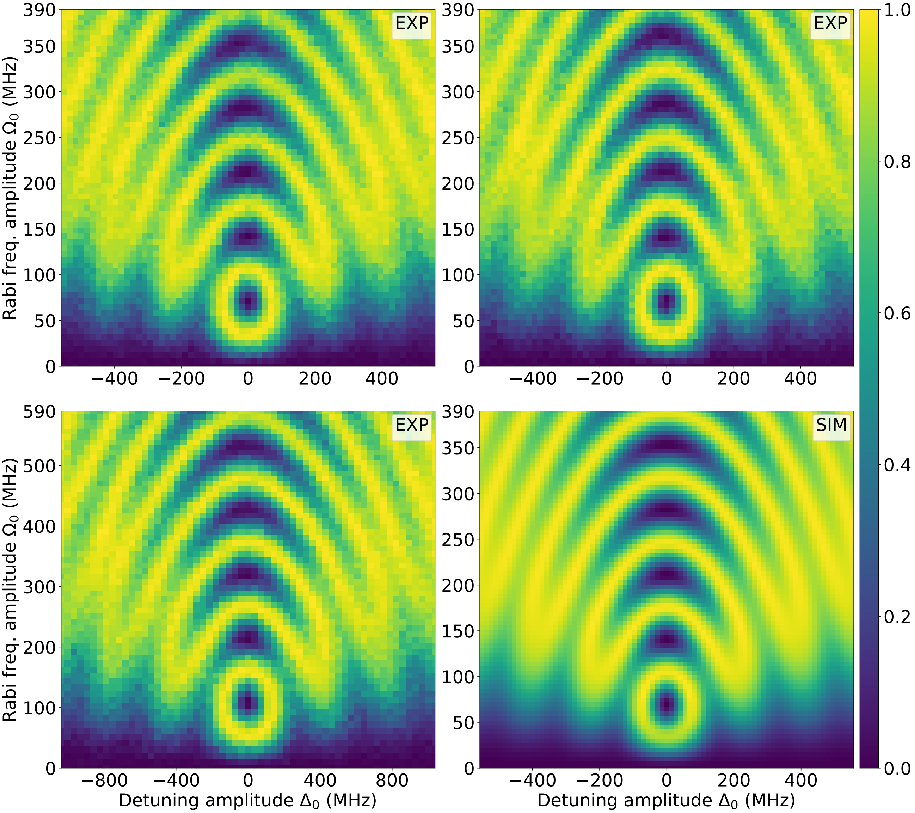} 
\end{tabular}
    \caption{[Colour online] 
    \textit{Top}: Rabi frequency (RF) $f(t)$ and detuning $g(t)$ of the LMSZ class of models with duration $T=88.9$~ns, top left: the first pair with constant RF ($\tau=28.3$~ns), top right: with cosine RF ($\tau=28.3$~ns), bottom left: with hyperbolic-secant RF ($\tau=22.2$~ns) and bottom right: simulation with constant RF ($\tau=28.3$~ns).
    \textit{Bottom}: The corresponding excitation landscapes (three experimental landscapes and a numerical simulation in the bottom right).\label{fig:lz}}
\end{figure}

The four two-dimensional colour maps in figure~\ref{fig:lz} show the measured transition probability of the finite LMSZ class of models in terms of the Rabi frequency amplitude $\Omega_0$ and the detuning amplitude $\Delta_0$.
The first three demonstrations (except the simulated landscape on bottom right) were performed using the three distinct LMSZ-class models that we just derived. 
The top left panel corresponds to model with the constant Rabi frequency, the top right panel shows the model with the cosine-shaped Rabi frequency, and the bottom left panel represents the model with the hyperbolic-secant-shaped Rabi frequency.
The bottom right plot is simulated numerically for the constant RF pair, effectively reproducing the model in the top left.
The detuning was emulated by using a time-dependent phase on the Rabi frequency, which is explained in section \ref{sec:phase_mod}. 
The first two pairs have $\tau=28.3$~ns and $T=88.9$~ns and the last pair has $\tau=22.2$~ns and $T=88.9$~ns.
The detailed specifications for the \textit{ibm\_kyiv} quantum processor can be found in section~\ref{sec-specs}.

The four landscapes in figure~\ref{fig:lz} are nearly identical, featuring chirp-symmetric arch-shaped fringes, typical for the LMSZ model.
The mean squared errors between the three experimental maps are $1.7\times10^{-3}$ between pairs 1 and 4, $0.8\times10^{-3}$ between pairs 4 and 8 and $1.4 \times10^{-3}$ between pairs 1 and 8 (procedure shown in section~\ref{sec-similarity}).
This equivalence shows that the same post-pulse transition probability pattern can emerge from multiple distinct combinations of Rabi frequency and detuning, further confirmed by the numerical simulation. 

\begin{table*}[tb]
\begingroup
\renewcommand{\arraystretch}{1.3}
\begin{tabular}{|c|c|c|c|c|c|c|}
\hline
\multirow{1}{*}
& Rabi &  & \multicolumn{2}{|c|}{LMSZ class} & \multicolumn{2}{|c|}{AEH class} \\ \cline{4-7} 
& frequency & Delos-Thorson & Detuning & Phase & Detuning & Phase  \\ 
& shape $f(x)$ & variable $s(x)$ & shape $g(x)$ & $\varphi(t)/(\Delta_0\tau)$ & shape $g(x)$ & $\varphi(t)/(\Delta_0\tau)$ \\ 
\hline 
1 & 1 & $x$ & $x$ & $\frac12 x^2$ & $\tan x$ & $-\ln\cos x$ \\
2 & $\frac{12}{\pi^2} x^2$ & $\frac{4}{\pi^2} x^3$ & $\frac{48}{\pi^4} x^5$ & $\frac{8}{\pi^4} x^6$ & $\frac{12}{\pi^2} x^2 \tan\frac{4}{\pi^2} x^3$ & $-\ln\cos\frac{4}{\pi^2} x^3$ \\ 
3 & $\frac{80}{\pi^4} x^4$ & $\frac{16}{\pi^4} x^5$ & $\frac{1440}{\pi^8} x^9$ & $\frac{144}{\pi^8} x^{10}$ & $\frac{80}{\pi^4} x^4 \tan\frac{16}{\pi^4} x^5$ & $-\ln\cos \frac{16}{\pi^4} x^5$\\ 
4 & $\frac{\pi}{2}\cos x$ & $\frac{\pi}{2}\sin x$ & $\frac{\pi^2}{8}\sin 2x$ & $\frac{\pi^2}{8}\sin^2 x$ & $f(x)\tan s(x)$ & $-\ln\cos s(x)$\\
5 & $2\cos^2 x$ & $x+\frac12\sin 2x $ & $f(x) s(x)$ & $\frac12 s(x)^2$ & $f(x)\tan s(x)$ & $-\ln\cos s(x)$\\ 
6 & $\frac{3\pi}{4}\cos^3 x$ & $\frac{\pi}{16} (9\sin x + \sin 3x)$ & $f(x) s(x)$ & $\frac12 s(x)^2$ & $f(x)\tan s(x)$ & $-\ln\cos s(x)$\\ 
7 & $\frac{8}{3}\cos^4 x$ & $x + \frac23\sin 2x + \frac{1}{12}\sin 4x$ & $f(x) s(x)$ & $\frac12 s(x)^2$ & $f(x)\tan s(x)$ & $-\ln\cos s(x)$\\ 
8 & $\sech x$ & $\arctan (\sinh x)$ & $f(x) s(x)$ & $\frac12 s(x)^2$ & $\tanh x$ & $\ln\cosh x$\\ 
9 & $\frac{\pi}{2}\sech^2 x$ & $\frac{\pi}{2}\tanh x$ & $f(x) s(x)$ & $\frac12 s(x)^2$ & $f(x)\tan s(x)$ & $-\ln\cos s(x)$\\ 
10 & $2\sech^3 x$ & $\arctan (\sinh x) + $\qquad\qquad & $f(x) s(x)$ & $\frac12 s(x)^2$ & $f(x)\tan s(x)$ & $-\ln\cos s(x)$\\ 
 &  & \quad\qquad$\sech x \tanh x$ & & & & \\ 
11 & $\frac{3\pi}{4}\sech^4 x$ & $\frac{\pi}{4}(2+\cosh x)\sech^2 x \tanh x$ & $f(x) s(x)$ & $\frac12 s(x)^2$ & $f(x)\tan s(x)$ & $-\ln\cos s(x)$\\ 
12 & $\dfrac{1}{1+x^2}$ & $\arctan x$ & $f(x) s(x)$ & $\frac12 s(x)^2$ & $\dfrac{x}{1+x^2}$ & $\frac12 \ln (1+x^2)$\\ 
13 & $\dfrac{2}{(1+x^2)^2}$ & $\arctan x + \dfrac{x}{1+x^2}$ & $f(x) s(x)$ & $\frac12 s(x)^2$ & $f(x)\tan s(x)$ & $-\ln\cos s(x)$\\ 
14 & $\dfrac{8}{3(1+x^2)^3}$ & $3\arctan x + \dfrac{x (5+3x^2)}{(1+x^2)^2}$ & $f(x) s(x)$ & $\frac12 s(x)^2$ & $f(x)\tan s(x)$ & $-\ln\cos s(x)$\\ 
15 & $\dfrac{16}{5(1+x^2)^4}$ & $15\arctan x +$\qquad\qquad & $f(x) s(x)$ & $\frac12 s(x)^2$ & $f(x)\tan s(x)$ & $-\ln\cos s(x)$\\ 
 &  & \quad\qquad$\dfrac{x(33+40x^2+15x^4)}{(1+x^2)^3}$ & & & & \\ 

16 & $\sqrt{\pi}\exp(-x^2)$ & $\frac{\pi}{2}\, \text{erf}\,x $ & $f(x) s(x)$ & $\frac12 s(x)^2$ & $f(x)\tan s(x)$ & $-\ln\cos s(x)$\\
\hline
\end{tabular}
\endgroup
\caption{Examples of specific models belonging to the Landau-Majorana-St\"uckelberg-Zener and Allen-Eberly-Hioe classes of models with $x=t/\tau$.
Note that $\int_L f(x) dx = \pi$, where $L$ is the time duration, for all pulse shapes.
For some models, the detuning shape $g(x)$ and the phase $\varphi(x)$ are given explicitly, whereas for others one need to replace $f(x)$ and $s(x)$ for the corresponding model.
The time duration of the Rabi frequency and detuning pair depends on the model --- it is $(-\frac12\pi,\frac12\pi)$ for models 1-7 and $(-\infty,\infty)$ for models 8-16.}
\label{tab-models}
\end{table*}

\subsection{AEH class of models}
The Allen-Eberly-Hioe model is defined by the pair
\be
\Omega(t) = \Omega_0\, \operatorname{sech}x,\quad \Delta(t) = \Delta_0 \tanh x.
\ee
The pulse area is $A=\pi\Omega_0\tau$.
Then 
$
s = \arctan (\sinh x)
$
and
$
x = \sinh^{-1}(\tan s).
$
Hence
$\Omega(t(s)) = \Omega_0 \cos s$,
$\Delta(t(s)) = \Delta_0 \sin s$,
and therefore
\be\label{Theta_AE}
\Theta(s) = \frac{\Delta_0}{\Omega_0} \tan s.
\ee
This is the St\"uckelberg variable for the AEH model.
The phase $\varphi$ in the variable $s(x)$ reads
\be\label{varphi_AE}
\varphi(s) = -\Delta_0 \tau \ln (\cos s).
\ee

The Allen-Eberly-Hioe model is analytically solvable.
Its transition probability is given by
\begin{align}
    P_{\text{AEH}}
    &= 1 - \frac{\cos^2 {\left(\pi\sqrt{\alpha^2 - \beta^2}\right)}}{\cosh^2 \left(\pi\beta\right)},
\label{eq-aeh-prob}
\end{align}
where $\alpha=\Omega_0 \tau/2$ and $\beta=\Delta_0\tau/2$, first derived in~\cite{Allen1975} and later in~\cite{Hioe1981}.

\begin{figure}
\begin{tabular}{c}
    \includegraphics[width=0.85\columnwidth]{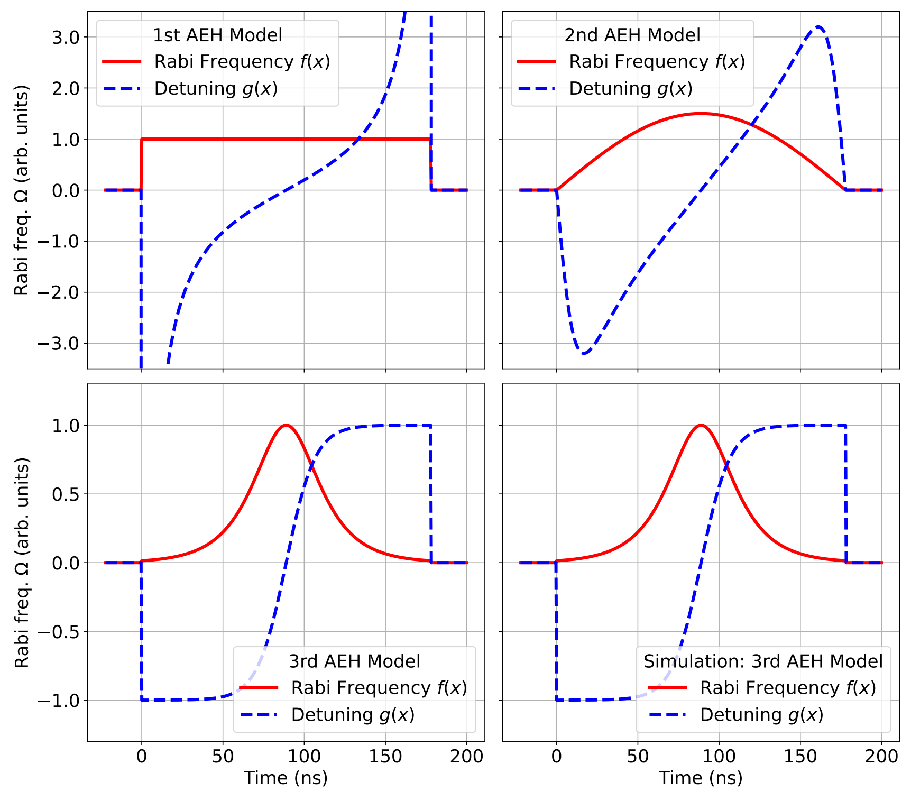} \\ 
    \hspace*{0.8cm}\includegraphics[width=0.85\columnwidth]{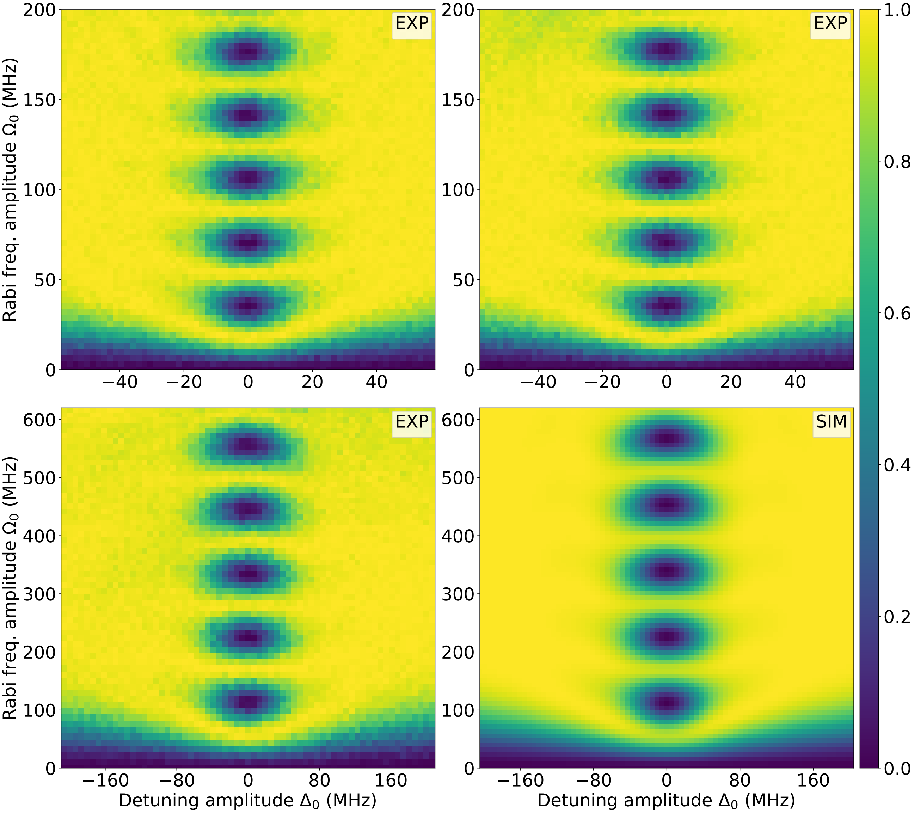}
\end{tabular}
    \caption{[Colour online] 
    \textit{Top}: Rabi frequency (RF) $f(t)$ and detuning $g(t)$ of the AEH class of models with duration $T=177.8$~ns, top left: the first pair with constant RF ($\tau=56.6$~ns), top right: with cosine RF ($\tau=56.6$~ns), bottom left: with hyperbolic-secant RF ($\tau=17.8$~ns) and bottom right: simulation with hyperbolic-secant RF ($\tau=17.8$~ns).
    \textit{Bottom}: The corresponding excitation landscapes (numerical simulation in the bottom right).\label{fig:ae}}
\end{figure}

Now let us assume that $\Omega(t)$ is the rectangular pulse of equation~\eqref{LMSZ model}. 
Then $s(x) = x$.
We replace this variable in equation~\eqref{eq:Delta} and find the corresponding detuning $\Delta(x) = \Delta_0 \tan x$. 

Consider now a Rabi frequency $\Omega(t)=\Omega_0 \frac\pi2 \cos x$, where $x=t/\tau \in [-\pi/2,\pi/2]$.
We have $s(x) = \frac\pi2 \sin x $.
Replacing this variable in equation~\eqref{eq:Delta}, we fix the detuning 
$\Delta(x) = \Delta_0 \left[\frac\pi2 \cos x \tan\left(\frac\pi2 \sin x\right)\right]$.

These and other pairs $\{\Omega(t),\Delta(t)\}$ belonging to the AEH class of models are presented in table~\ref{tab-models}.
One can also find a pair of the AEH family with a linear detuning, which
requires the Rabi frequency
\be
\Omega(t) = \frac{\Omega_0 |t|}{\sqrt{e^{t^2/\tau^2}-1}} , \quad 
\Delta(t) = \Delta_0 t.
\ee

Several Rabi frequency/detuning pairs of the Allen-Eberly-Hioe class were also validated on the IBM Quantum processor. 
Figure~\ref{fig:ae} displays the excitation landscapes of pairs 1, 4 and 8 presented in table~\ref{tab-models} in top left, top right and bottom left respectively.
A numerical simulation of the transition probability in the hyperbolic-secant RF model that models the properties of the transmon system can be found in the bottom right plot.
A notable aspect of this model are the prominent off-resonant patches where complete population transfer occurs.
These can be seen coloured in yellow in the panels of figure~\ref{fig:ae}.
All four excitation patterns are consistent, particularly in the central elliptical regions where the transition probability drops to zero.
The MSE between pairs No. 1 and 4 is $0.5\times10^{-3}$.
The mean squared error between pairs No. 1 and 8 is higher at $1.6\times10^{-3}$ and is comparable to the one between pairs No. 4 and 8 at $1.5\times 10^{-3}$.
In our demonstration, the first two Allen-Eberly models (top row) were applied with $\tau=56.6$~ns and $T=177.8$~ns, while the third model (bottom left) was performed with $\tau=17.8$~ns and $T=177.8$~ns.

\begin{figure}[h!]
    \centering
    \includegraphics[width=0.8\textwidth]{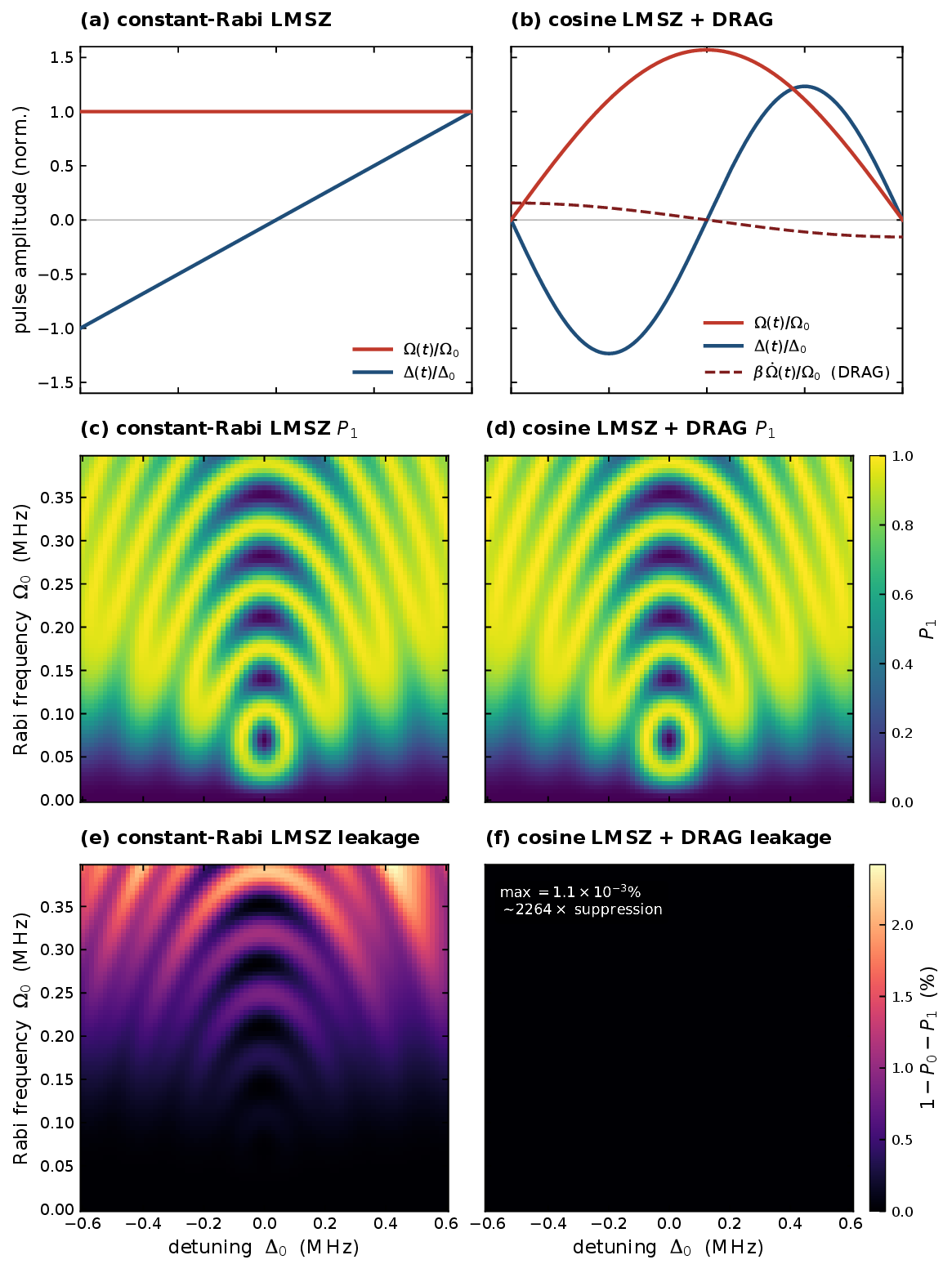}
    \caption{Leakage suppression via isoprobability shape switching within the LMSZ class. 
    (a) Pulse shapes of the canonical finite LMSZ model: constant Rabi frequency $\Omega(t) = \Omega_0$ (red) and linearly swept detuning $\Delta(t) \propto t/\tau$ (blue). The constant envelope makes $\dot{\Omega} = 0$ everywhere, so the DRAG correction $\beta\dot{\Omega}$ vanishes identically and DRAG cannot be applied. 
    (b) Pulse shapes of the cosine-Rabi isoprobability sibling (model 2 of Table I): $\Omega(t) = \frac{\pi}{2}\Omega_0\cos(t/\tau)$ (red solid) and the corresponding Delos-Thorson detuning $\Delta(t) = \frac{\pi^2}{8}\Delta_0\sin(2t/\tau)$ (blue solid). The smooth envelope admits a nonzero DRAG quadrature $\beta\dot{\Omega}(t)$ (red dashed), here with $\beta = 0.1$. 
    (c),(d) Numerically simulated transition probability $P_1$ as a function of the Rabi frequency amplitude $\Omega_0$ and chirp parameter $\Delta_0$ for the constant-Rabi model and the cosine-Rabi model with DRAG, respectively. The two landscapes are nearly identical, confirming that the isoprobability property is preserved under shape switching. 
    (e),(f) Total leakage $1 - P_0 - P_1$ out of the computational subspace $\{|0\rangle, |1\rangle\}$ for the same two models, displayed on the same color scale. Without DRAG the constant-Rabi pulse accumulates up to $2.4\%$ leakage at high drive amplitudes. Switching to the cosine-Rabi sibling and applying DRAG reduces this to below $1.1 \times 10^{-3}\%$, a suppression of more than three orders of magnitude across the entire landscape. Simulations use an 8-level transmon Hamiltonian with parameters matching qubit 14 of the \texttt{ibm\_kyiv} processor ($E_J/h = 71.1\,\text{GHz}$, $E_C/h = 1.66\,\text{GHz}$, $\omega_d/2\pi = 4.604\,\text{GHz}$), pulse duration $T = 88.9\,\text{ns}$, and time scale $\tau = 28.3\,\text{ns}$.}
    \label{fig:lmsz_drag}
    
\end{figure}

\section{DRAG-based LMSZ -- absolute leakage suppression}

In the DRAG construction, the leading-order leakage to $\ket{2}$ is suppressed by an out-of-phase quadrature $Q(t) = -\dot{\Omega}(t)/\Delta_{\text{anh}}$, whose role is to cancel the non-adiabatic contribution of the leakage state generated by the time-varying Rabi drive~\cite{Motzoi2009, Gambetta2011}. 
The correction is proportional to $\dot{\Omega}$ -- it is designed to target the envelope's rate of change. 
For a textbook LMSZ pulse with constant Rabi frequency, this term collapses to zero throughout the pulse interval -- there is no derivative for DRAG to remove. 
The isoprobability framework resolves this asymmetry by trading the flat envelope for a smooth twin within the same LMSZ class -- specifically, the cosine-Rabi member (No.~4 in table~\ref{tab-models}) -- while preserving the post-pulse transition probability.

We numerically demonstrate the resulting leakage suppression in figure~\ref{fig:lmsz_drag}, using the first eight levels of a transmon Hamiltonian fitted to qubit 14 of the \texttt{ibm\_kyiv} processor (appendix~\ref{sec-specs}). 
Panels (a) and (b) show the two pulse pairs under comparison: the original constant-Rabi LMSZ model with linearly swept detuning (a), and its smooth cosine-Rabi twin $\Omega(t) = (\pi/2)\Omega_0 \cos(t/\tau)$ paired with the Delos--Thorson detuning $\Delta(t) = (\pi^2/8)\Delta_0 \sin(2t/\tau)$, augmented by a DRAG quadrature $\beta\dot\Omega(t)$ with $\beta = 0.1$ (b). 
The constant envelope of (a) leaves no derivative for DRAG to act on, whereas the smooth envelope of (b) does.

Panels (c) and (d) present the transition probability $P_1$ across a $41 \times 41$ grid spanning the $(\Delta_0, \Omega_0)$ plane. 
The two landscapes are visually indistinguishable: both reproduce the chirp-symmetric arch-shaped fringes characteristic of the LMSZ class, with maxima reaching $P_1 = 0.9975$ for the constant-Rabi pulse and $P_1 = 0.9998$ for the cosine-Rabi DRAG-corrected twin. This near-equality numerically confirms that isoprobability shape switching together with the DRAG correction preserves the full LMSZ transition pattern, complementing the experimental validation of Section~\ref{sec:isoprobability-twins}.

The contrast between the two pulses appears in panels (e) and (f), which display the total leakage $1 - P_0 - P_1$ on a common color scale. 
For the constant-Rabi LMSZ pulse, leakage out of the computational subspace reaches up to $2.4\%$ at large drive amplitudes, with a mean leakage of $0.53\%$ across the landscape. 
For the cosine-Rabi DRAG-corrected twin, the same scale renders the leakage essentially invisible: the peak leakage is $1.1 \times 10^{-3}\%$ and the mean leakage is $2.8 \times 10^{-4}\%$. 
This represents a peak suppression factor of $\sim\!2200$ and a mean suppression factor of $\sim\!1900$ -- more than three orders of magnitude across the entire $(\Delta_0, \Omega_0)$ landscape.

The combined picture is that the canonical constant-Rabi LMSZ pulse and its smooth cosine-Rabi twin produce indistinguishable transition probabilities (panels c, d), while differing by more than three orders of magnitude in their leakage to non-computational states (panels e, f). 
The isoprobability framework thus reconciles the analytic tractability of the LMSZ model with the standard DRAG leakage-suppression toolkit, a combination that is structurally impossible for the constant-Rabi model itself.

\section{Discussion}


The principal advantage of the framework is that it decouples the choice of pulse shape from the choice of analytical model. 
The same post-pulse population transfer can be reached through a constant Rabi envelope, a cosine envelope, a hyperbolic-secant envelope, or any other physically convenient choice, with the corresponding detuning fixed by the Delos--Thorson construction. 
This new degree of freedom allows the pulse shape to be chosen to match hardware constraints rather than to dictate them. 
Most strikingly, we numerically showed that switching from the canonical constant-Rabi LMSZ pulse to its smooth cosine-Rabi twin makes the model compatible with DRAG, suppressing leakage to non-computational states by more than three orders of magnitude (figure~\ref{fig:lmsz_drag}) while leaving the LMSZ transition landscape essentially unchanged. 
The combination of an analytic transition probability with the standard DRAG leakage-suppression toolkit is structurally impossible for the canonical LMSZ pulse itself due to the vanishing derivative of $\Omega$.

The practical scope of the framework is broadest on platforms where the Rabi envelope is directly shapeable, which includes essentially all microwave- and laser-driven qubit architectures. 
In Rydberg atom physics, chirped-pulse adiabatic rapid passage routinely excites selected Rydberg states through avoided crossings, and the dense ladder of nearby Rydberg states makes leakage suppression an important concern.
This makes smooth LMSZ siblings combined with derivative-based corrections a natural fit~\cite{Beterov2011}. 
Trapped-ion motional-mode control, where chirped sideband pulses act on a Fock-state ladder, falls into the same category, as does cold-molecule state preparation across closely-spaced vibrational and rotational levels. 
In NMR and MRI, where AEH-derived sech/tanh pulses are already the key to $B_1$-insensitive inversion~\cite{Silver1984, Garwood2001}, the framework supplies an analytic recipe for migrating between shape variants while preserving the inversion fidelity exactly.

A complementary regime is provided by fixed-gap LZSM platforms --- fluxonium and composite-transmon qubits~\cite{Zhang2021, Campbell2020} --- in which the transverse coupling is set by an avoided-crossing gap and only the longitudinal detuning can be shaped through a flux or baseband pulse. 
There, switching the Rabi envelope is unavailable, but the framework still applies in the opposite direction: the constant-coupling member of the AEH class --- $\Omega = \Omega_0$ paired with a detuning $\Delta(t) \propto \tan(t/\tau)$ on $|t/\tau| < \pi/2$ --- shows that a fixed-coupling architecture can, in principle, realise an AEH-class transition without ever leaving its native control regime. 
More generally, every isoprobability class admits a constant-Rabi representative whose detuning is forced by the class generator, yielding a menu of analytical chirp shapes for hardware where one can only control its longitudinal bias.

Beyond these specific scenarios, the framework provides a generic route for adapting analytically solvable models to platform-specific constraints: shorter pulse durations, gentler edges to mitigate truncation artefacts, smaller local gradients to avoid AWG discretisation errors, and compatibility with the wider toolkit of derivative-based leakage corrections beyond DRAG. 
We expect this flexibility to be useful wherever an exact post-pulse transition probability is required but the canonical pulse shape of the class cannot be cleanly implemented.

\section{Conclusion}

In conclusion, we presented the concept of isoprobability models --- models with different Rabi frequency and detuning shapes that reach the same post-pulse transition probability distribution along different evolution paths.
We found and compiled 16 different models for each of two different classes --- the LMSZ and AEH classes --- in table~\ref{tab-models} by developing on the Delos-Thorson equivalence principle, which grants us the ability to construct an infinite number of these sibling models.

We used IBM's 127-qubit \textit{ibm\_kyiv} transmon-based processor to measure and validate the transition probability of six members of the aforementioned classes. 
Our demonstration and subsequent MSE calculation confirmed the equivalence between the post-pulse transition probabilities of three members of the LMSZ and AEH classes.
This was performed despite a key hardware constraint of the system --- lack of direct time-dependent control of the detuning --- by modulating the Rabi frequency's phase instead of the detuning.

A highly impactful application of this methodology, it enables applying the DRAG technique to the Landau-Zener model. 
This DRAG-corrected LMSZ model eliminates leakage, reducing it by more than three orders of magnitude, all this while boosting the transition probability to the excited state.
These are only part of the applications of this analytically-derived pulse shape equivalence, since shape switch can bear other benefits, such as extending models to other previously limited hardware.

\ack{
We gratefully acknowledge the Karoll Knowledge Foundation for providing financial support to I.S.M. during the preparation of this manuscript. 
Their contribution was invaluable, not only to this work but in encouraging the author's ongoing scientific endeavours. 

This research is supported by the Bulgarian national plan for recovery and resilience, Contract No. BG-RRP-2.004-0008-C01 (SUMMIT), Project No. 3.1.4, 
and by the European Union's Horizon Europe research and innovation program under Grant Agreement No. 101046968 (BRISQ). 
We acknowledge the use of IBM Quantum services and the supercomputing cluster PhysOn at Sofia University for this work. 
The views expressed are those of the authors and do not reflect the official policy or position of IBM or the IBM Quantum team.
}

\appendix

\section{Time-dependent phase control}\label{sec:phase_mod}
Experiments involving models with time-dependent detuning usually rely on full control of the excitation pulse, typically achieved by shaping both the detuning and the Rabi frequency.
However, in some systems, direct control of the detuning is not feasible, demanding an alternative approach.
In these cases, one may instead model the detuning by using a time-dependent phase of the driving field, which we relate to the detuning by
\be\label{phase}
\varphi(t) = \int_0^t \Delta(t')\,dt', \quad
\varphi(\sigma) = \int_0^\sigma \Theta(\sigma') d\sigma'.
\ee
effectively generating a variable detuning. 
This equivalence between phase and detuning can be shown by a transformation of the Hamiltonian from the Schr\"odinger picture to the interaction picture,

\be
\textbf{H}_{i}(t) = \tfrac{1}{2}\begin{bmatrix}
0 & \Omega(t)\,e^{i \varphi(t)}\\
\Omega^*(t)\,e^{-i \varphi(t)} & 0
\end{bmatrix},
\label{eq-ham-schro}
\ee
by using the population-preserving phase transformation
\be
\textbf{U}(t) = \begin{bmatrix}
e^{i \varphi(t) /2} & 0\\
0 & e^{-i \varphi(t) /2}
\end{bmatrix}.
\label{eq-U}
\ee

Thence, the Rabi frequency phase $\varphi(t)$ can be used to produce an effective time-dependent detuning $\Delta(t)$ in the two-state system.
We note that although using a time-dependent detuning and a time-dependent phase are mathematically equivalent, the respective physical implementations can be vastly different.

\section{Experimental specifications of IBM Quantum processors}\label{sec-specs}

The models were experimentally tested against measurements on qubit 14 of the 127-qubit transmon quantum processor -- IBM's \texttt{ibm\_kyiv}, an Eagle r3 Processor~\cite{IBM, Koch2007}.
Qubit 14 had an excitation frequency of $4.6040$~GHz and anharmonicity of $-0.3075$~GHz.

The measurements of the finite Landau-Majorana-St\"uckelberg-Zener~(LMSZ) and Allen-Eberly-Hioe~(AEH) and  class models were recorded on 29 December 2024.
The $T_1$ decoherence time was 387.50~\textmu s, while the $T_2$ was 376.31~\textmu s.
All experimental data points in the 2D excitation landscapes of figures~\ref{fig:lz} and~\ref{fig:ae} were recorded over 200 shots.

We employed the Qiskit Pulse framework for Python to access and control the low-level configurations of the quantum processor, enabling the implementation of desired pulse shapes. 
While the framework is freely accessible, it comes with certain inherent limitations:

(i) unfeasibility of drive with time-dependent detuning;

(ii) a soft cap on the total pulse duration, typically in the range of several microseconds;

(iii) a maximum amplitude constraint, restricted to 1 in Qiskit's arbitrary units;

(iv) a minimum pulse duration imposed by discretisation, set at $2/9$~ns, equal to 1~\texttt{dt}, the elementary unit of time, used in IBM's quantum systems;

(v) basis measurements limited to the first two energy levels, meaning any leakage is aggregated with the population of the excited state.

Despite these constraints, the framework offers a broad array of resources for constructing gates, methods and algorithms. 
With creative solutions, these limitations can often be circumvented, acting as more of a challenge than a fundamental barrier.

\section{Alignment and MSE-based similarity of 2D maps}\label{sec-similarity}

We conduct thorough examination of the equivalence between the three experimental 2D transition-probability maps within each class, ensuring consistency of the models within each family. 
The procedure compares a pair of 2D maps $P_1(\Delta,\Omega)$ and $P_2(\Delta,\Omega)$ by conducting offsets and trims on all sides while looking to minimise the mean squared error (MSE) between them.
The alignment procedure is motivated by difference in axes and resolutions of the experimental 2D maps and aims to quantify agreement after removing trivial misregistrations, deriving a single minimised value for the MSE.

First, each map is resampled by bilinear interpolation onto a common uniform grid
\begin{equation}    
    \{\tilde{\Delta}_i\}_{i=1}^{N_\Delta},\qquad
    \{\tilde{\Omega}_j\}_{j=1}^{N_\Omega},
\end{equation}
yielding arrays $\tilde{P}_k(i,j)=P_k(\tilde{\Delta}_i,\tilde{\Omega}_j)$ for $k\in\{1,2\}$. 
The resampling aims to drastically increase the resolution and thus hinder discretisation errors.

To account for residual misalignment and edge effects, we optimise over integer pixel shifts of the maps $(\Delta x,\Delta y)$ applied to $\tilde{P}_2$ and over non-negative trim margins
\begin{equation}
c^{(k)}_{x,L},\;c^{(k)}_{x,R},\;c^{(k)}_{y,B},\;c^{(k)}_{y,T}\quad (k=1,2),
\end{equation}
which clip the left/right and bottom/top edges of each map. For a parameter vector
\begin{equation}
    \theta=\big(\Delta x,\Delta y, c^{(1)}_{x,L},c^{(1)}_{x,R},c^{(1)}_{y,B},c^{(1)}_{y,T}, c^{(2)}_{x,L},c^{(2)}_{x,R},c^{(2)}_{y,B},c^{(2)}_{y,T}\big),
\end{equation}
we extract the maximal common overlap region $\mathcal{R}(\theta)$ after shifting $\tilde{P}_2$, trimming both maps and resampling the overlap so that both arrays are shape-matched.

We minimise the MSE over the overlap:
\begin{equation}
\operatorname{MSE}(\theta)=\frac{1}{|\mathcal{R}(\theta)|}\sum_{(i,j)\in\mathcal{R}(\theta)}
\big(\tilde{P}_1(i,j)-\tilde{P}_2^{(\theta)}(i,j)\big)^2,
\end{equation}
where $\tilde{P}_2^{(\theta)}$ denotes the shifted and clipped $\tilde{P}_2$. A derivative-free Powell optimisation over bounded integer shifts and trims yields the optimiser $\hat\theta$. 
We report both the unaligned error,
\begin{equation}
\operatorname{MSE}_{\text{pre}}=\frac{1}{N_\Delta N_\Omega}\sum_{i,j}\big(\tilde{P}_1(i,j)-\tilde{P}_2(i,j)\big)^2,
\end{equation}
and the aligned error $\operatorname{MSE}_{\text{post}}=\operatorname{MSE}(\hat\theta)$.

On our data, the maps were resampled to a size of (10000, 10000), while the pixel shifts and clips were bounded up to 500 pixels or 5\% of the entire plots to ensure only small modifications.
The alignment reduced the error by approximately one order of magnitude, i.e.,
\begin{equation}
\operatorname{MSE}_{\text{post}} \ll \operatorname{MSE}_{\text{pre}} \quad (\sim\!4-12\times\ \text{improvement}).
\end{equation}
For human validation we also inspected the difference maps $\tilde{P}_2^{(\hat\theta)}-\tilde{P}_1$ on $\mathcal{R}(\hat\theta)$.

\roles{
I.S.M. contributed to the formal analysis, investigation, software, validation and visualisation, whereas
N.V.V. contributed to the conceptualisation, funding acquisition, project administration, resources and supervision. 
The methodology and the writing of the original draft were joint contributions.}

\data{The research data that is presented in this publication is available from the \textsc{qiskit\_experiments} repository located at \url{https://github.com/ivo53/qiskit\_experiments/tree/master/data/kyiv/14/power\_broadening\%20(narrowing)}.
It can be found in the \textsc{lz1\_pulses}, \textsc{lz4\_pulses}, \textsc{lz8\_pulses}, \textsc{ae1\_pulses}, \textsc{ae4\_pulses} and \textsc{ae8\_pulses} folders, the newest data from 30 December.}

\end{document}